# A Scalable, Linear-Time Dynamic Cutoff Algorithm for Molecular Dynamics

Paul Springer, Ahmed E. Ismail, and Paolo Bientinesi

Aachen Institute for Advanced Study in Computational Engineering Science,
RWTH Aachen University,
Schinkelstr. 2, 52062 Aachen, Germany
{springer,ismail,pauldj}@aices.rwth-aachen.de
http://hpac.rwth-aachen.de/

**Abstract.** Recent results on supercomputers show that beyond 65K cores, the efficiency of molecular dynamics simulations of interfacial systems decreases significantly. In this paper, we introduce a dynamic cutoff method (DCM) for interfacial systems of arbitrarily large size. The idea consists in adopting a cutoff-based method in which the cutoff is chosen on a particle-by-particle basis, according to the distance from the interface. Computationally, the challenge is shifted from the long-range solvers to the detection of the interfaces and to the computation of the particle-interface distances. For these tasks, we present linear-time algorithms that do not rely on global communication patterns. As a result, the DCM algorithm is suited for large systems of particles and massively parallel computers. To demonstrate its potential, we integrated DCM into the LAMMPS open-source molecular dynamics package, and simulated large liquid/vapor systems on two supercomputers: SuperMuc and JUQUEEN. In all cases, the accuracy of DCM is comparable to the traditional particle-particle particle-mesh (PPPM) algorithm, while the performance is considerably superior for large numbers of particles. For JUQUEEN, we provide timings for simulations running on the full system (458, 752 cores), and show nearly perfect strong and weak scaling.

## 1 Introduction

Molecular dynamics (MD) is a vital tool for computational chemistry, materials science, biophysics, and many other fields. The basic idea underpinning MD is the direct numerical integration of Newton's laws of motion, which require the frequent evaluation of forces between atomistic- or molecular-scale "particles". Although the underlying model is conceptually simple, significant challenges arise because of the enormous number of particles found even in nanoscopic systems. In this paper, we discuss the development implementation, and parallelization of a new force computation algorithm especially designed for systems consisting of large number of particles which demand the equivalent of "Tier-0" computing resources.

Practically, MD calculations are limited by available computational resources, with typical simulations today involving anywhere from $10^4$ to $10^7$ particles, although simulations of $10^9$ atoms or more have been reported in the literature [24,26]. In general, larger simulations are preferable to smaller ones because smaller simulations can be affected by finite-size effects that reduce the accuracy of the calculations by introducing spurious correlations between particles [11]. Moreover, in principle, every particle can interact with every other particle, making the inherent complexity of MD $\mathcal{O}(N^2)$. Thus, a primary driver of active research in MD is reducing the algorithmic complexity of the force calculations while preserving both accuracy and scalability.

To integrate Newton's equations of motion, one needs to calculate the interaction forces among all of the particles in the system. Formally, these forces can be calculated using any scheme that correctly accounts for all forces present. Calculations are typically divided into *bonded* and *non-bonded* forces:

$$\mathbf{F} = \mathbf{F}_{\text{bonded}} + \mathbf{F}_{\text{non-bonded}}, \tag{1}$$

where bonded forces result from the topological structure of molecules, while non-bonded forces account for all other interactions (such as gravity and electromagnetic effects). Since the calculation of bonded forces already has linear complexity, our focus is on the non-bonded forces, which can have complexity up to $\mathcal{O}(N^2)$. In particular, we focus on a class of forces known as *dispersion* forces, which represent forces that exist as a result of the gravitational interaction between particles, independent of any other internal and external forces in the system. These dispersion forces are typically calculated as a sum of pairwise interactions between particles, with the strength of the interaction depending on the distance between them:

$$\mathbf{F}_{\text{disp}} = \sum_{i<j} \mathbf{F}(r_{ij}), \tag{2}$$

where $r_{ij} = |\mathbf{r}_i - \mathbf{r}_j|$ is the distance between atoms $i$ and $j$.

Until recently, dispersion forces were treated using a cutoff on the distance, beyond which they were assumed to be negligible:

$$\mathbf{F}(r_{ij}) = \begin{cases} \mathbf{F}(r_i j), & r_{ij} \leq r_c \\ 0 & r_{ij} > r_c \end{cases}, \tag{3}$$

where $r_c$ is the user-specified "cutoff" parameter. This approach, also referred to as a short-range method, reduces the complexity of the force calculation in Eq. 2 from $\mathcal{O}(N^2)$ to $\mathcal{O}(Nr_c^3)$, where $N$ is the number of particles. Such cutoff-based methods are sufficiently accurate for homogeneous systems, whose composition is uniform throughout the simulation volume. However, in heterogeneous systems, with nonuniform spatial density that leads to the existence of interfaces, assuming isotropic behavior can cause major technical problems, as illustrated in Fig. 1a. The red particle in the right-hand circle indicates the behavior in the "bulk" part of the system, where isotropic behavior can be assumed and the errors introduced by Eq. 3 largely cancel, as can be seen from the green particles around the red particle, whose force contributions essentially negate one another. However, near the interface, such as the blue particle on the left, this cancellation of errors is impossible, as the distribution of atoms across the interface is far from isotropic. This breakdown, which can lead to completely inaccurate results, has been demonstrated by a number of different researchers [2, 6, 10, 17]. Successful resolution of this problem is critical in a range of applications, including industrial uses such as spreading and coating of films [14] as well as modeling of the dynamics of cell membranes [4].

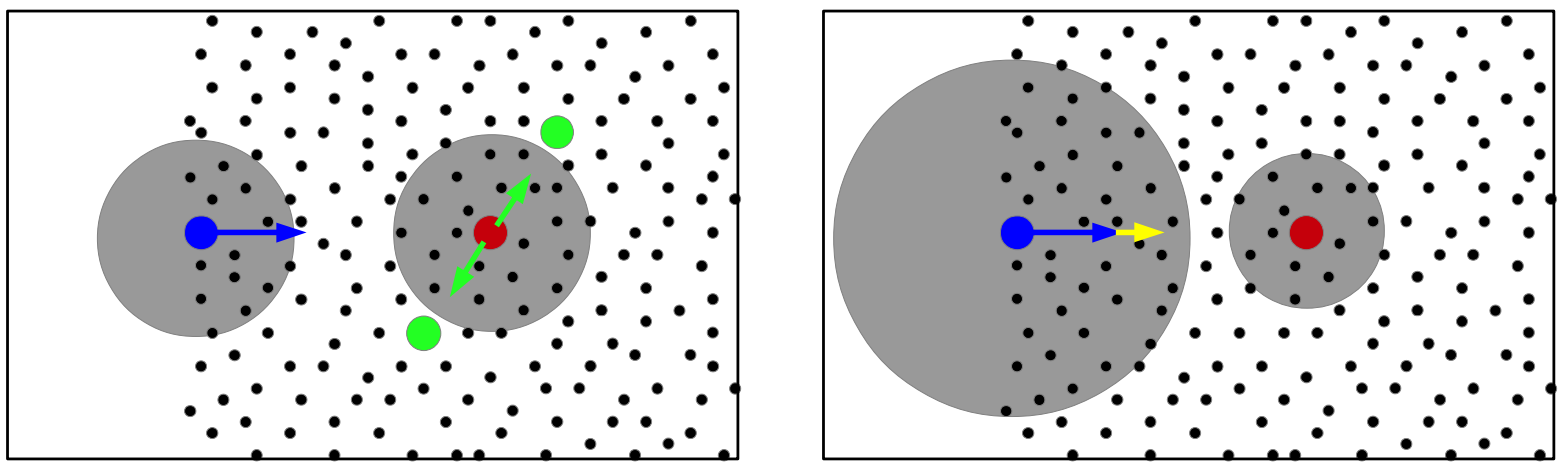

(a) Static cutoff. (b) Dynamic cutoff.

Fig. 1: Interfacial system using a (a) static cutoff and a (b) dynamic cutoff. Gray area denotes the cutoff. The arrows represent the force acting on a particle. The yellow arrow depicts the additional force contribution due to a larger cutoff.

A naïve solution to account for the "missing" interactions in Fig. 1b would be to increase the magnitude of the cutoff $r_c$; doubling and even tripling the magnitude of the cutoff has been proposed [33]. Such an approach is inherently undesirable, as the $\mathcal{O}(Nr_c^3)$ complexity of the method means that doubling the cutoff leads to an eight-fold increase in the cost of the pairwise computations. In response, a number of so-called "long-range solvers" have been developed to reduce the overall complexity. Most of these approaches are based on Ewald summation methods [8], which rely on Fourier transforms to reduce the complexity of the force calculations to as little as $\mathcal{O}(N \log N)$, with implementations based on the classical method [13] and mesh-based approaches such as the particle-particle particle-mesh (PPPM) [15, 16] and particle mesh Ewald (PME) [30] methods. Other approaches, such as the multilevel summation al-gorithm [27] further reduce the complexity to $\mathcal{O}(N)$. However, these methods all suffer from a critical drawback: they require global communications between different processors. Consequently, their scalability eventually decreases as the number of cores increase and the cost of all-to-all communications becomes prohibitively expensive [15], as shown in Fig. 2. Although Sun et al. [26] were able to optimize communications within the PME solver in NAMD, an open-source MD package, to achieve good strong scaling for a system with $10^8$ particles on up to 298 992 cores, their approach still requires all-to-all communications and has a complexity of $\mathcal{O}(N \log N)$.

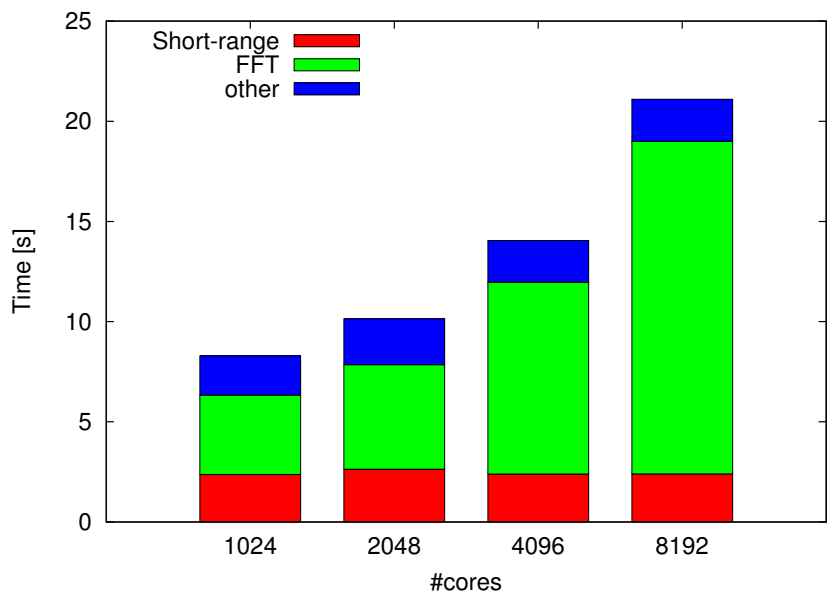

Fig. 2: Weak scaling of the PPPM long range solver (from the LAMMPS framework) for an interfacial system with 1200 particles per core (IBM BG/Q).

Since short-range methods exhibit errors at the interface that are typically two orders of magnitude larger than the errors for particles in the bulk phase [15], it would be helpful to direct the computational resources to where they are most needed. The approach we introduce in this paper, which we call the *dynamic cutoff method* (DCM), circumvents the need for all-to-all communications by making the cutoff a particle-dependent property. As shown in Fig. 1b, particles located in bulk regions, where the isotropic assumption is valid, can be handled with a small cutoff, while particles close to the interface are assigned a larger cutoff. Consequently, computational demands are kept to a minimum while maintaining high accuracy. The DCM is closely related to static cutoff methods [12, 28] and, as a result, inherits their good properties, such as strictly local communication and good scalability. To make DCM competitive with state-of-the-art solvers, we have also developed a fast and scalable algorithm to detect interfaces. A similar method involving adaptive cutoffs, using a derived error estimate rather than the relative location of the particles to determine the cutoff, was recently proposed [29]. However, as that algorithm still relies on the use of fast Fourier transforms, its large-scale scalability remains questionable.

This paper outlines the development of the dynamic cutoff method and the associated interface detection method, which has been parallelized and extended to three dimensions. These algorithms were incorporated into the open-source LAMMPS package [9, 20], one of the most widely used MD simulators currently available. We show that our implementation of the dynamic cutoff algorithm achieves linear-time scaling for interfacial systems, even when utilizing the entire JUQUEEN supercomputer at the Forschungszentrum Jülich.

## 2 Dynamic Cutoff Method

The core idea of the DCM is to circumvent the use of long-range solvers by adaptively choosing the cutoff on a particle-by-particle basis, using small cutoffs for particles far away from the interface and increasingly larger cutoffs as one approaches the interface. Clearly, this strategy requires knowledge of the position and the time evolution of the interface.

The computational tasks involved in one iteration of the DCM are shown schematically in Fig. 3. First, the interface is identified (box 1); for each particle, the distance from the interface is computed, and the cutoff for each atom is determined and assigned (box 2). Like classical short-range methods, the DCM builds a neighbor list (box 3), enabling each particle to access its neighbors in $\mathcal{O}(1)$ time. Finally, pairwise forces are computed (box 4), and the positions and velocities of all particles are updated (box 5). At this point, the next iteration begins. Since in typical MD simulations the interface changes very slowly, the interface detection and neighbor-list build need not be executed every iteration.

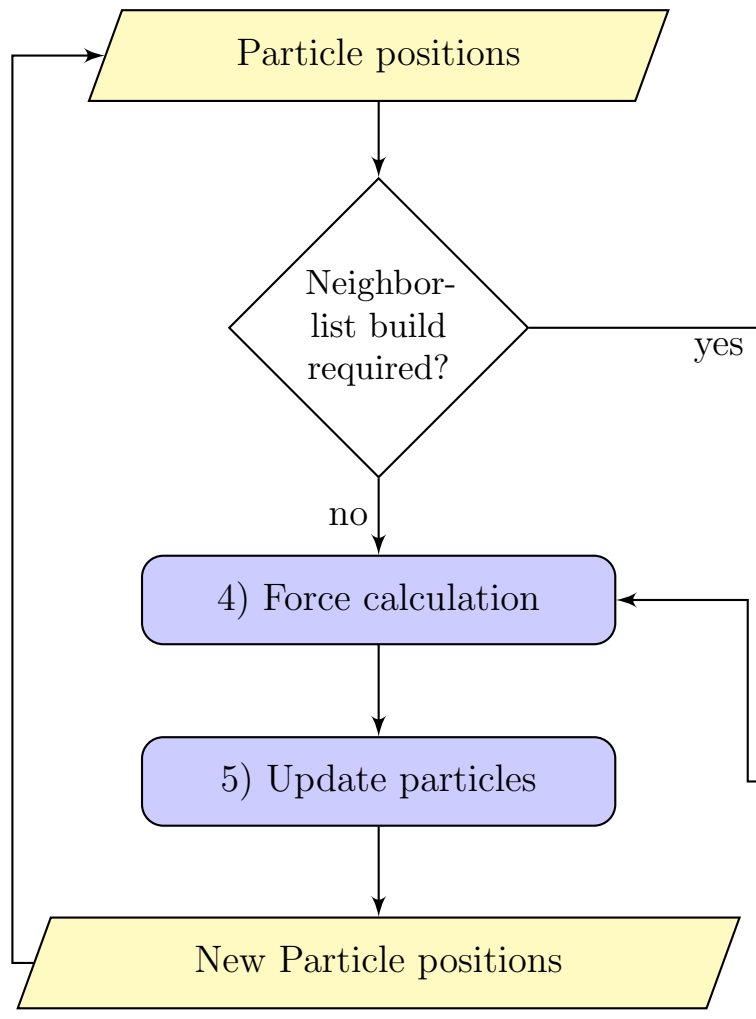

Fig. 3: Schematic overview of the dynamic cutoff method.

Before describing the individual tasks, we briefly discuss our overall parallelization strategy for distributed memory environments. Following the main scheme used by LAMMPS, the physical domain is spatially decomposed and assigned to the underlying three-dimensional grid of MPI processes. Each process $p$ is responsible for one subdomain $L^p \subseteq L$ of the computational domain $L = L_x \times L_y \times L_z \subset R^3$, and has an average memory requirement of $\mathcal{O}(N/P)$, where $N$ and $P$ are the total number of particles and processes, respectively. Particles can migrate between pairs of neighboring processes [20]. As we will show, irrespective of the task performed, each process only communicates with its direct neighbors, so that global communication patterns are entirely avoided.[1] DCM exhibits a linear dependence on the number of particles in the system.

The first four tasks enumerated in Fig. 3 are covered in detail in the next subsections. The final task, responsible for the updates of the particle positions and velocities, uses velocity Verlet integration [28], the standard integration scheme in MD simulations, and is therefore not discussed further.

[1] With the exception of a reduction operation to identify the maximum of a scalar in the interface detection method.

### 2.1 Interface Detection

Our approach for a fast interface detection is inspired by algorithms for binary image segmentation. The main idea is that an interface delineates the regions of the physical domain in which the density of particles changes; with this in mind, we treat particle densities as gray-scale values and apply image segmentation techniques to the data. In three dimensions, this effectively becomes a gray-scale volume of voxels (3D pixel). As shown in Fig. 4a, to create the gray-scale volume from the particle positions, all particles are binned into small 3D $h \times h \times h$ "boxes",[2] effectively decomposing each subdomain $L^p$ into small 3D boxes $b_{x,y,z} \subset L^p$; this operation only requires neighbor-neighbor communication.

At this stage, each box is treated as a voxel and is assigned a gray-scale value according to its relative particle density (Fig. 4b). Based on this gray-scale volume, the segmentation (Fig. 4c) can be computed as the minimization of the piecewise constant Mumford-Shah functional for two-phase segmentation [1, 18]. The result is a binary classification of the boxes, differentiating high-density phases (e.g., liquid) from low-density ones (e.g., vapor). A distributed-memory implementation of this third stage boils down to the parallelization of a 3D finite-difference stencil [3, 9]. Starting from the Mumford-Shah algorithm from the QuocMesh open-source library [21], an accurate 3D segmentation algorithm for shared-memory architectures, we developed an MPI-based parallelization, adding support for the periodic boundary conditions typically used in MD simulations; this algorithm is now included in QuocMesh.

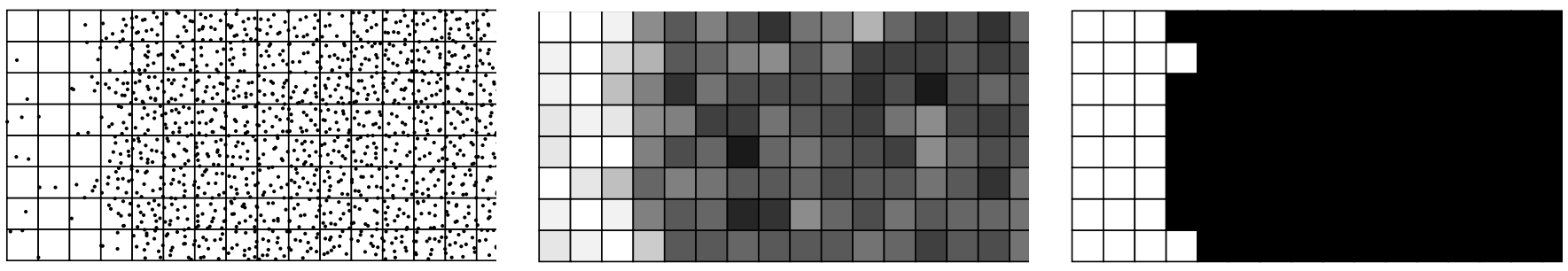

(a) Binning (b) Gray-scale volume (c) Segmented volume

Fig. 4: Interface detection: a 2D slice from a 3D domain.

The output of the three stages shown in Fig. 4 is a segmented volume $S$:

$$S = \{s_{x,y,z} \in \{0, 1\} \mid 0 \leq x < N_x, 0 \leq y < N_y, 0 \leq z < N_z\}, \qquad (4)$$

where $N_x, N_y$ and $N_z$ are the number of local boxes on a given processor; $s_{x,y,z}$ equals 0 if the corresponding box $b_{x,y,z}$ belongs to the low-density phase, and $s_{x,y,z} = 1$ otherwise. The interface is then determined by adjacent boxes with discontinuous values (Fig. 4c). The minimization of the Mumford-Shah functional might result in the set $S$ presenting low-density "bubbles" inside the high-density region and vice versa. Depending on their size, such bubbles can be interpreted as false-detections and cause a noticeable performance degradation (bubbles yield additional interfaces and hence unnecessarily large cutoffs). For this reason, we apply a parallel multi-stage filtering algorithm to identify and remove connected components smaller than a user-specified volume. This step, which again requires communication only between neighboring processes, is described in [25]. In contrast to other interface-detection methods [5, 19, 22], our algorithm is so fast that it does not affect the performance of an MD simulation.

[2] The edge length $h$ determines the resolution of the interface and can be automatically chosen at the beginning of the simulation.

### 2.2 Cutoff assignment

The objective of this task is to adaptively assign a suitable cutoff to each individual particle; to this end, the set $D^p$ of box-interface distances,[3] and from this the particle-interface distances $\delta$ are then derived. Two numerical methods for approximating the box-interface distances are the Fast Marching Method (FMM) [23] and the Fast Sweeping Method (FSM) [32]. Let $N_v$ be the number of voxels of the system; FMM has a complexity $\mathcal{O}(N_v \log N_v)$ and is generally more accurate than FSM. However, since FSM has a preferred complexity of $\mathcal{O}(N_v)$ and in practice its accuracy is sufficient for the DCM, we adopted an FSM-based approach.

Because the cutoffs for particles vary only for particles within a given distance from the interface, we need not compute the exact distance between the interface and each box. Instead, it suffices to carry out the calculations up to a certain threshold distance $r_c^{\text{grid}}$, since beyond this distance, the "minimum cutoff" will be applied, regardless of the actual distance from the interface. This problem formulation makes it possible to devise a fast sweeping method that only requires local communication and the reduction of a scalar.

Zhao et al. proposed two parallel algorithms/implementations for the FSM [31]. Since scalability is one of our main concerns, we developed our own version of FSM which was specifically tailored to the needs of our problem. Henceforth we call our implementation a cutoff-based fast sweeping method (CFSM).

As shown in Algorithm 2.1, the CFSM propagates box-interface distances outwards until the distance is larger than the threshold $r_c^{\text{grid}}$.[4] Visually, the algorithm unfolds as a wave that starts at the interface and flows outwards until it has traveled the maximum distance $r_c^{\text{grid}}$. The set of box-interface distances $D^p$, local to process $p$, is initialized such that boxes at the interface and in the low-density region are assigned a distance of 0, while boxes in the high-density region are assigned a distance of $+\infty$ (line 1). All boxes adjacent to the interface and have nonzero distance are added to queue $Q$, which keeps track of the remaining boxes to be processed. After initialization, $Q$ is processed breadth-first. For each box, indexed as $(x, y, z)$, the distance to the interface is computed using a FSM on the local subdomain (line 4), the distance of box $b_{x,y,z}$ is updated (line 8),

[3] $D^p = \{d_{x,y,z} \in \mathbb{R} \,|\, 0 \leq x < N_x, 0 \leq y < N_y, 0 \leq z < N_z\}$; the superscript $p$ indicates that this set is computed on each process, in parallel.

[4] The exact value for the threshold is not important here. More information is provided in [25].

**Algorithm 2.1** Distributed cutoff-based fast sweeping method.

1: initialize($D^p$, $Q$) ▷ Add interfacial boxes to $Q$<br>
2: **for** $0 \leq \text{iter} < \text{iter}^{\max}$ **do**<br>
3:     **for all** boxes $(x, y, z) \in Q$ **do**<br>
4:         $d^{new} \leftarrow$ solveDistance($D^p$, $(x, y, z)$) ▷ local FSM<br>
5:         **if** $d^{new} \leq r_c^{\text{grid}}$ **then**<br>
6:             **if** $|d_{x,y,z} - d^{new}| > \Delta e$ **then**<br>
7:                 $\Delta e \leftarrow |d_{x,y,z} - d^{new}|$<br>
8:             $d_{x,y,z} \leftarrow d^{new}$<br>
9:             addNeighborsToQueue($\widetilde{Q}$, $(x, y, z)$)<br>
10:     swapQueues($Q$, $\widetilde{Q}$)<br>
11:     emptyQueue($\widetilde{Q}$)<br>
12:     $\Delta e \leftarrow$ MPI_Allreduce($\Delta e$, MAXIMUM)<br>
13:     **if** $\Delta e < \epsilon$ **then**<br>
14:         **break**<br>
15:     ghostExchange($D^p$) ▷ Local communication<br>
16:     addModifiedBoundariesToQueue($D^p$, $Q$)

and all its adjacent boxes are added to the auxiliary queue $\widetilde{Q}$ (line 9). Once all boxes in $Q$ have been processed, the queues $Q$ and $\widetilde{Q}$ are swapped (line 10), and the maximum difference $\Delta e$ between the old and new distances (line 7) is reduced among all processes (line 12). If $\Delta e < \epsilon$ for some threshold $\epsilon$ (line 13), all processes terminate; otherwise, each process communicates its boundary boxes to its neighbors (line 15), adds the received boundary to $Q$ (line 16), and enters a new iteration. Typically, two to four iterations suffice for convergence. For completeness, we point out that for any box within the threshold $r_c^{\text{grid}}$, this implementation yields the same results as the algorithms proposed by Zhao et al. [31].

From the set of computed box-interface distances $D^p$, the particle–interface distances $\delta_i \in \mathbb{R}$, $i \in [1, \ldots, N]$, can be estimated via trilinear interpolation. The cutoff $r_c$ of each particle is then chosen as a function of $\delta_i$.[5] In all cases, particles at the interface or within the low-density phase are assigned a larger cutoff, up to $r_c^{\max}$, than particles further away; beyond a given distance from the interface, particles in the high-density phase are assigned the minimum cutoff $r_c^{\min}$.

### 2.3 Neighbor-list Build

Neighbor lists in MD simulations allow particles to access all of their neighbors in constant time. Hockney et al. [12] introduced the *linked-cell* method, which bins particles into cells of edge $r_c$, thereby restricting the search for neighbors of particle $i$ to the cell containing particle $i$ and its 26 neighbors. An alternative technique, introduced by Verlet et al. [28], uses a neighbor (or Verlet) list for each particle $i$: the indices of all particles that interact with particle $i$ are stored

[5] Possible interpolation functions and the resulting accuracy are discussed in [25].

(i.e., $r_{ij} \leq r_c$) (Algorithm 2.2). In practice, a skin distance $r_s > 0$ is introduced, so that particles with $r_{ij} \leq r_c + r_s$ are stored; this allows reuse of the neighbor list over multiple timesteps. A drawback of this technique is that the neighbor list must be updated frequently [7]. Currently, the most common approach combines these techniques and bins all particles only when a neighbor-list build is required.

Algorithm 2.2 outlines the steps needed to build a neighbor list in the specific context of the DCM; it is assumed that all particles are already spatially sorted into bins. For each particle $i$, the algorithm loops over all particles $j$ in the neighboring bins *jBin* and adds $j$ to the neighbor list of particle $i$ if $r_{ij}$ is less than the cutoff $r_c[i]$ of atom $i$.

**Algorithm 2.2** Neighbor-list build.

1: **for all** local atoms $i$ **do**
2:     $\text{nNbrs}[i] \leftarrow 0$
3:     $\text{iBin} \leftarrow \text{getBin}(i)$
4:     **for all** $\text{jBin} \in \text{neighbors}(\text{iBin})$ **do**
5:         **for all** atoms $j$ of jBin **do**
6:             $\mathbf{r_{ij}} \leftarrow \mathbf{r_i} - \mathbf{r_j}$
7:             **if** $|\mathbf{r_{ij}}| < r_c[i]$ **and** $i \neq j$ **:**
8:                 $\text{nbrs}_i[\text{nNbrs}[i]] \leftarrow j$
9:                 $\text{nNbrs}[i] \leftarrow \text{nNbrs}[i] + 1$

**Algorithm 2.3** Force calculation.

1: **for all** atoms $i$ **do**:
2:     $\mathbf{f_i} \leftarrow 0$
3:     **for all** neighbors $k$ of $i$ **do**
4:         $j \leftarrow \text{nbrs}_i[k]$
5:         $\mathbf{r_{ij}} \leftarrow \mathbf{r_i} - \mathbf{r_j}$
6:         **if** $|\mathbf{r_{ij}}| < r_c[i]$ **:**
7:             $f \leftarrow \text{forceLJ}(|\mathbf{r_{ij}}|)$
8:             $\mathbf{f_i} \leftarrow \mathbf{f_i} - f \times \mathbf{r_{ij}}$
9:     $\text{force}[i] \leftarrow \text{force}[i] + \mathbf{f_i}$

The varying cutoffs of DCM pose additional challenges for efficient implementation of the neighbor-list build. With a static cutoff, all particles traverse the same stencil of neighboring cells. If applied to the DCM, this static approach would result in poor performance because particles with a small cutoff traverse the same volume as particles with the maximum cutoff. For instance, assuming $r_c^{min} = \frac{1}{2} r_c^{max}$ and a typical skin distance $r_s = 0.1 r_c^{max}$, all particles would traverse a volume $V_{cube} = (3(r_c^{max} + r_s)^3)$ to find their neighbors. However, particles assigned the minimum cutoff have their neighbors within the much smaller volume $V_{min} = \frac{4}{3}\pi {r_c^{min}}^3$. Thus, only $V_{min}/V_{cube} \approx 1.5\%$ of all particle-particle calculations would contribute to the neighbor-list build (i.e., Line 7 of Algorithm 2.2 would return `false` 98.5% of the time). Since the neighbor-list build is memory-bound, this approach would nullify any performance benefit gained using dynamic cutoffs.

The solution lies in the choice of the bins' edge length: instead of $l^{\max} = r_s + r_c^{\max}$, we use an edge length of $l^{\min} = r_s + r_c^{\min}$ or smaller. While this results in having to traverse a slightly larger stencil of neighboring cells, the traversed volume is considerably smaller than $V_{cube}$ and results in many fewer spurious distance calculations. Note that the complexity of this improved neighbor-list build is hidden in Line 4 of Algorithm 2.2. This optimization yields a 4×–6× speedup of the neighbor-list build over the binning with edge length $l^{\max}$.

### 2.4 Force Calculation

Compared to classical short-range methods, the force calculations within the DCM (Algorithm 2.3) show two striking differences: a particle-dependent cutoff (shown in red), and the inapplicability of Newton's third law of motion[6] (henceforth called N3). Since the cutoff is independently assigned to each particle, the fact that particle $j$ is "influenced" by particle $i$ does not imply that particle $i$ is influenced by particle $j$. Effectively, neglecting N3 means that the update $\mathbf{f_j} \leftarrow \mathbf{f_j} + f \times \mathbf{r_{ij}}$, which would appear in Algorithm 2.3 right after Line 8, is not performed. Computationally, this results in twice as many force calculations, but also allows a better memory access pattern, since costly scattered memory accesses for $\mathbf{f_j}$ are avoided.[7]

Our DCM implementation is based on the existing short-range Lennard-Jones solver in LAMMPS. We developed both a pure MPI implementation, as well as a hybrid MPI + OpenMP-based shared-memory parallelization that allows us to start multiple threads per MPI rank, reducing both the memory requirements and communication overhead. While the shared-memory implementation consists of simple OpenMP directives—for instance, (**#pragma omp for schedule**(dynamic,20)) before the outermost loops of Algorithm 2.2 and 2.3 suffices to distribute the loops across multiple threads—we stress that the default static schedule would result in severe load imbalance (due to different cutoffs for different particles). By contrast, a dynamic schedule, using tasks of about 20 particles, results in almost perfectly load-balanced simulations. The benefits of dynamic scheduling become more apparent as more threads are involved. Simulations with 256 MPI ranks and 16 threads per rank[8] show a speedup of 1.4× for the neighbor list and 2.2× for the force-calculation kernels over static scheduling.

## 3 Simulation Methodology

We present performance results for two interfacial systems: one with a planar interface (Fig. 5a) and another with a non-planar interface (Fig. 5b). As the processor count increases, the area of the interface in the planar system is scaled proportionally in two dimensions (i.e., creating a large plane), and in the non-planar system is extended along its cylindrical axis.

Both accuracy and performance are compared to the particle-particle particle-mesh solver (PPPM), a state-of-the-art long-range algorithm included in the LAMMPS package. Specifically, the measurements for the static cutoff method and PPPM are obtained using LAMMPS, version *30Oct14*, with the OpenMP user-package installed. For all experiments, the settings for PPPM are chosen according to [16] and are considered to be optimal. Unless otherwise specified, the minimum and maximum cutoff of DCM are respectively set to $r_c^{min} = 3.0$ and $r_c^{max} = 8.0$, such that the resulting accuracy is comparable to that of PPPM. The experiments were carried out on two different supercomputing architectures: SuperMuc and JUQUEEN.

[6] If a body $i$ exerts a force $f$ onto another body $j$, then $j$ exerts a force $-f$ on $i$.
[7] This is why most GPU implementations of force calculations also neglect N3.
[8] Running on 1024 cores on the BlueGene/Q supercomputer with simultaneous multi-threading enabled for four threads per core.

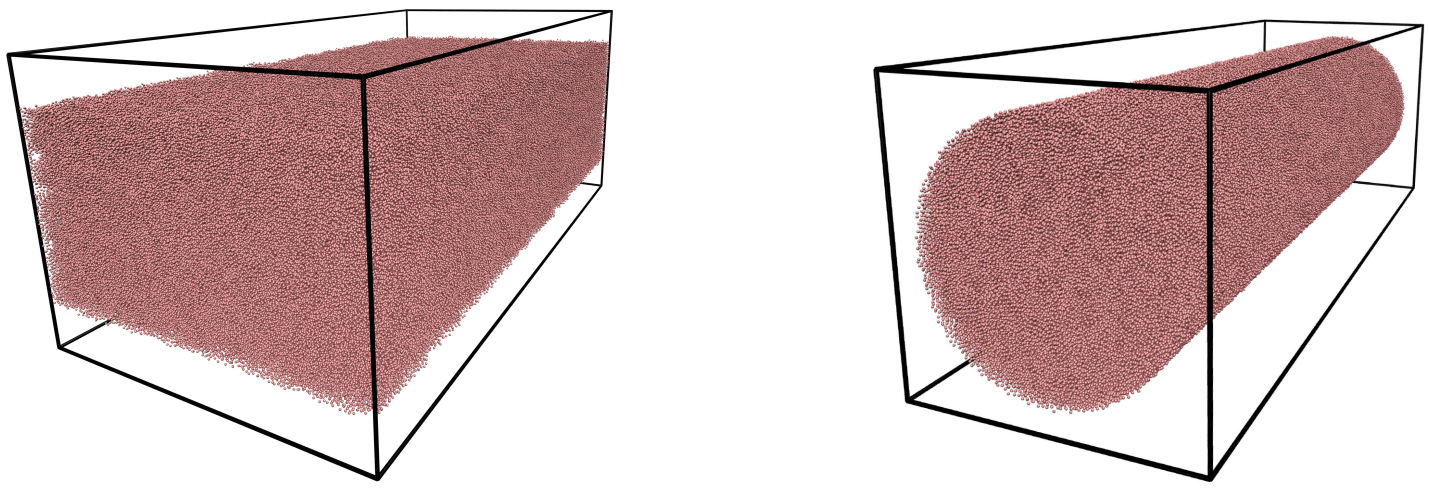

(a) Planar interface. (b) Non-planar interface.

Fig. 5: The two interfacial systems used in this publication.

*SuperMUC.* The SuperMUC supercomputer at the Leibniz Supercomputing Centre is based on Intel's Sandy Bridge architecture, with 147.456 cores and 288 TB of main memory arranged in 18 "islands" of 512 nodes each. A node consists of two Intel Xeon E5-2680 CPUs with a total of 16 cores. We used Intel's C++ compiler *icpc 14.0.3* with compiler flags *-O3 -restrict -ip -unroll0 -openmp -xAVX*.

*JUQUEEN.* The JUQUEEN supercomputer at Forschungszentrum Jülich is a IBM Blue Gene/Q machine with 28.672 nodes organized in 28 racks, each comprising 1024 nodes. A node consists of 16 IBM PowerPC A2 cores, and 16GBs of DDR3 memory, for a total of 458.752 cores and 448 TB of main memory. We used IBM's C compiler *xlc++ 12.01* with compiler flags *-O3 -qarch=qp -qtune=qp -qsmp=omp -qsimd=auto -qhot=level=2 -qprefetch -qunroll=yes*.

## 4 Performance and Accuracy Results

### 4.1 Accuracy

As discussed in Section 1, every physical property (e.g., pressure, density) in a molecular dynamics simulation relies on accurate force calculations. We therefore choose to measure the per-particle error in the forces perpendicular to the interface (in these experiments, along the $z$-direction) to validate the correctness of DCM. A detailed accuracy analysis of DCM is beyond the scope of this paper. The error $\Delta f_{i_z}$ of particle $i$ is computed as follows:

$$\Delta f_{i_z} = f^*_{i_z} - f_{i_z} \tag{5}$$

where $f^*_{i_z}$ denotes the correct force for particle $i$ along the $z$-direction.[9] We only show the component of the error perpendicular to the interface because this is much larger than the error along either of the other directions.

Fig. 6 shows the error in the $z$-component of the forces for the *planar* system with 19.200 particles for PPPM and DCM with different maximum cutoffs. First, we note that the errors of DCM and PPPM are comparable in magnitude. Second, larger cutoffs for DCM lead to more accurate results, as expected. Third, in DCM the errors are smaller at the interface; this is critical, as the error strongly influences the physical behavior [15,16,25]. Finally, as the dashed lines indicate, one can see that our interface detection method correctly identifies the interface.

[9] $f^*_{i_z}$ is computed by the accurate (but expensive) Ewald long-range solver.

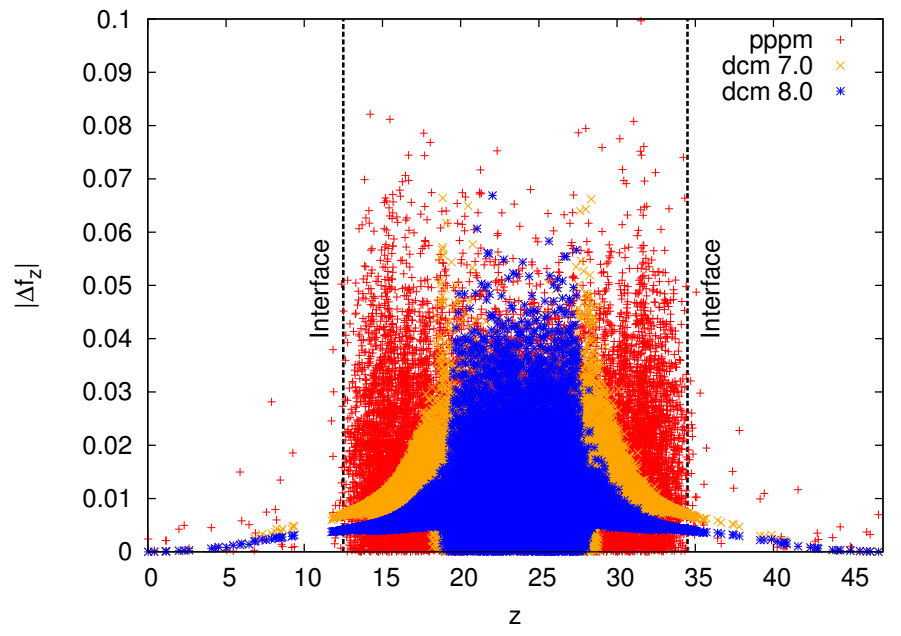

Fig. 6: Absolute error in the $z$-direction (i.e., perpendicular to the interface) for the planar system with $19,200$ particles. Each cross corresponds to a single particle. DCM uses a minimum cutoff of $r_c^{min} = 3.0\sigma$.

| DCM setting | 3.0/7.0 | 3.0/8.0 | 3.5/8.0 | 3.0/9.0 | 3.5/9.0 |
|---|---|---|---|---|---|
| Speedup | 2.31 | 2.66 | 2.38 | 2.42 | 2.23 |

Table 1: Speedup of DCM over a static cutoff with $r_c = r_c^{max}$ for the planar system with 12 million particles. The DCM setting reads as $r_c^{min}/r_c^{max}$. The experiments were run on JUQUEEN using 1024 cores (256 MPI ranks and 16 threads per rank).

### 4.2 Performance

We compare the performance of DCM with the static cutoff method and PPPM on two different architectures. Table 1 shows the speedup of DCM over its static counterpart for the planar system. The simulation was run on JUQUEEN, on 1024 cores with 1200 particles per core. The static cutoff is set to the maximum DCM cutoff: $r_c = r_c^{max}$. Despite not exploiting Newton's third law, the DCM outperforms the static cutoff version by at least a factor of 2.2. We note that these speedups are highly dependent on the ratio between the number of particles at the interface and away from the interface; depending on this ratio, even higher speedups can be expected by incorporating Newton's third law.

Fig. 7 presents the strong and weak scalability on SuperMUC for the non-planar system and on JUQUEEN for the planar system. On both architectures, the weak scaling experiments were performed with 1200 particles per core, while the strong scaling experiments use a system with roughly $4 \times 10^7$ particles.

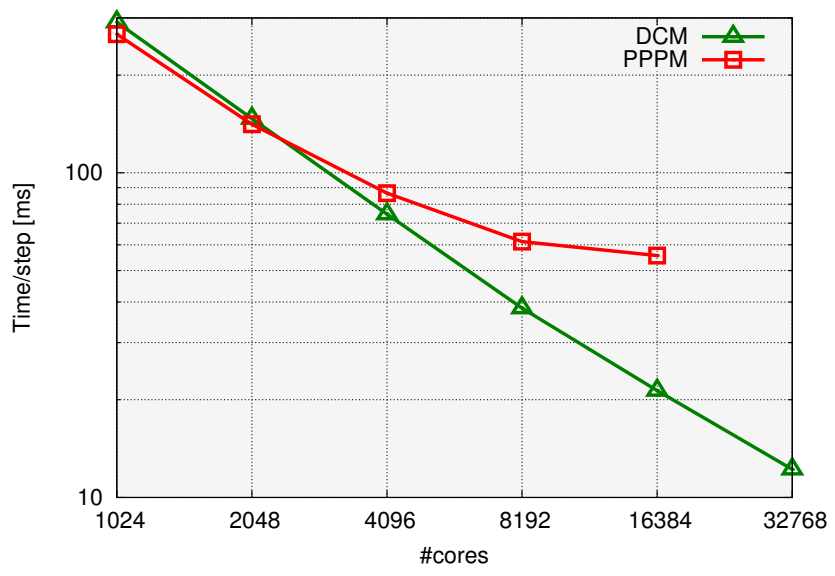

(a) SuperMUC: Strong scaling

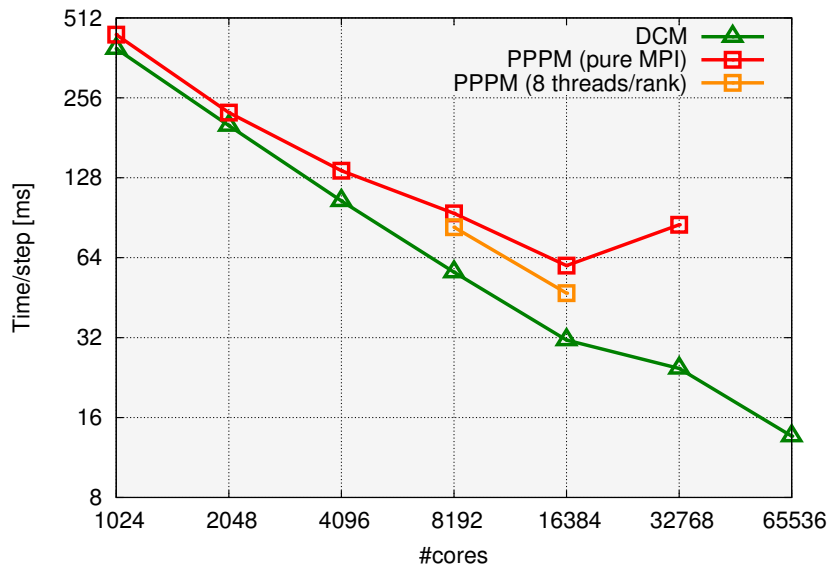

(b) JUQUEEN: Strong scaling

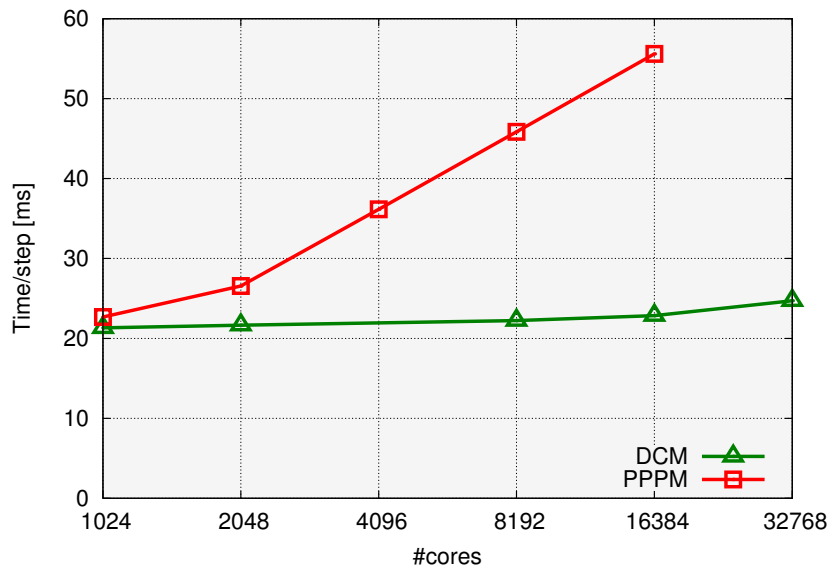

(c) SuperMUC: Weak scaling

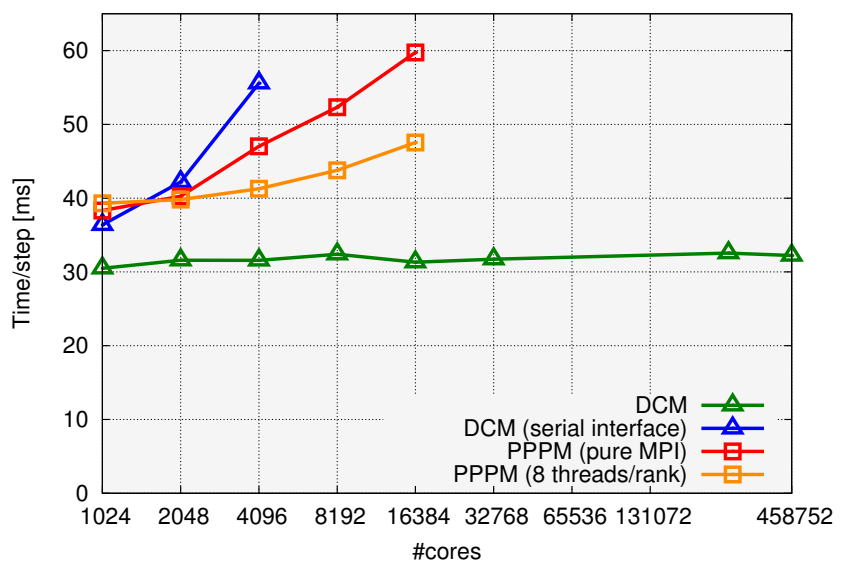

(d) JUQUEEN: Weak scaling

Fig. 7: Strong (a, b) and weak (c, d) scalability for PPPM and DCM.

Figs. 7a and 7b show a head-to-head comparison between DCM and PPPM in terms of strong scalability. The results for the pure MPI version of PPPM expose a degradation in scalability starting from 4k and 8k cores on SuperMUC and JUQUEEN, respectively. On JUQUEEN, we also ran an hybrid multithreaded+MPI version of PPPM, using 8 threads per MPI rank; this configuration attained somewhat better timings than the pure MPI version, but consistently crashed on 32k cores or more. The DCM achieved convincing (nearly-linear) scalability on both systems.

The weak scalability behaviour of DCM and PPPM is illustrated in Figs. 7c and 7d. On both systems, the trend of PPPM (red line) matches exactly the behaviour presented in 2: use of FFTs causes the scalability to progressively deteriorate as the number of cores increases. This phenomenon is only delayed in the hybrid multithreaded + MPI version of PPPM (orange line in Fig. 7d), which eventually follows the same diverging trend. The results for the DCM on JUQUEEN clearly indicate the need for a parallel interface detection method, since the serial implementation (blue line), although extremely fast, eventually becomes the bottleneck for the entire DCM. Finally, we direct our attention to the DCM (green line) with parallel interface detection (as described in Sec. 2.1): for both the planar and the nonplanar systems, scalability is nearly perfect. Indeed, Fig. 7d reveals that on JUQUEEN it was possible to scale the system up to $5.5 \times 10^8$ particles on all $458,752$ available cores, while attaining ideal weak scalability.

## 5 Conclusion

We have developed a dynamic cutoff method to study large-scale interfacial and heterogeneous systems in molecular dynamics simulations containing millions of particles on massively parallel supercomputers. Our method is based on making the cutoff for force calculations between particles a particle-dependent property. We have implemented DCM as part of the open-source LAMMPS MD package and showed that it exhibits desired properties such as (1) linear-time complexity, (2) local communication, and (3) ideal weak- and strong-scaling on up to $458,752$ cores. Moreover, our performance results show that DCM outperform state-of-the-art algorithms for large interfacial Lennard-Jones systems. These experiments suggest that DCM is a promising algorithm for massively parallel supercomputers.

We have also presented a scalable interface detection method for non-planar interfaces. This method is fast enough to be applicable in real time throughout the course of an MD simulation, which may open the door to a wide variety of new MD applications. This interface detection method enabled us to preserve the linear scaling of the DCM for short-ranged potentials.

Even though not investigated further in this paper, DCM can be used as a replacement for short-range calculations within mesh-based Ewald solvers (e.g., PPPM). This allows to shift computational workload from the FFTs to the short-range calculations and therefore should improve the scalability of these solvers as well.

**Acknowledgments.** The authors gratefully acknowledge financial support from the Deutsche Forschungsgemeinschaft (German Research Association) through grant GSC 111, computing resources on the supercomputer JUQUEEN at Jülich Supercomputing Centre (JSC) (project ID: e5430301) and the Gauss Centre for Supercomputing/Leibniz Supercomputing Centre (project ID: pr84za), and Edoardo Di Napoli and Benjamin Berkels for helpful discussions.

## References

1. B. Berkels. An unconstrained multiphase thresholding approach for image segmentation. In *Proceedings of the Second International Conference on Scale Space Methods and Variational Methods in Computer Vision (SSVM 2009)*, volume 5567 of *Lecture Notes in Computer Science*, pages 26–37. Springer, 2009.
2. E. Blokhuis, D. Bedeaux, C. Holcomb, and J. Zollweg. Tail corrections to the surface tension of a lennard-jones liquid-vapour interface. *Molecular Physics*, 85(3):665–669, 1995.
3. T. Bohlen. Parallel 3-d viscoelastic finite difference seismic modelling. *Computers & Geosciences*, 28(8):887–899, 2002.

4. R. Bradley and R. Radhakrishnan. Coarse-grained models for protein-cell membrane interactions. *Polymers*, 5(3):890–936, 2013.
5. F. Bresme, E. Chacón, and P. Tarazona. Molecular dynamics investigation of the intrinsic structure of water–fluid interfaces via the intrinsic sampling method. *Physical Chemistry Chemical Physics*, 10(32):4704–4715, 2008.
6. G. A. Chapela, G. Saville, S. M. Thompson, and J. S. Rowlinson. Computer simulation of a gas–liquid surface. part 1. *Journal of the Chemical Society, Faraday Transactions 2: Molecular and Chemical Physics*, 73(7):1133–1144, 1977.
7. A. A. Chialvo and P. G. Debenedetti. On the use of the Verlet neighbor list in molecular dynamics. *Computer Physics Communications*, 60(2):215–224, 1990.
8. P. Ewald. Die Berechnung optischer und elektrostatischer Gitterpotentiale. *Annalen der Physik*, 369:253–287, 1921.
9. M. Griebel, S. Knapek, and G. Zumbusch. *Numerical simulation in molecular dynamics*. Springer, 2007.
10. M. Guo, D.-Y. Peng, and B. C-Y Lu. On the long-range corrections to computer simulation results for the Lennard-Jones vapor-liquid interface. *Fluid phase equilibria*, 130(1):19–30, 1997.
11. T. L. Hill. *Thermodynamics of Small Systems*. Dover Publications, 2013.
12. R. Hockney, S. Goel, and J. Eastwood. Quiet high-resolution computer models of a plasma. *Journal of Computational Physics*, 14(2):148–158, 1974.
13. P. J. in 't Veld, A. E. Ismail, and G. S. Grest. Application of Ewald summations to long-range dispersion forces. *Journal of Chemical Physics*, 127:144711, 2007.
14. R. E. Isele-Holder and A. E. Ismail. Atomistic potentials for trisiloxane, alkyl ethoxylate, and perfluoroalkane-based surfactants with tip4p/2005 and application to simulations at the airwater interface. *The Journal of Physical Chemistry B*, 118(31):9284–9297, 2014.
15. R. E. Isele-Holder, W. Mitchell, J. R. Hammond, A. Kohlmeyer, and A. E. Ismail. Reconsidering Dispersion Potentials: Reduced Cutoffs in Mesh-Based Ewald Solvers Can Be Faster Than Truncation. *Journal of Chemical Theory and Computation*, 9(12):5412–5420, Dec. 2013.
16. R. E. Isele-Holder, W. Mitchell, and A. E. Ismail. Development and application of a particle-particle particle-mesh ewald method for dispersion interactions. *The Journal of Chemical Physics*, 137(17):174107, 2012.
17. A. E. Ismail, M. Tsige, P. J. in 't Veld, and G. S. Grest. Surface tension of normal and branched alkanes. *Molecular Physics*, 105(23-24):3155–3163, 2007.
18. D. Mumford and J. Shah. Optimal approximations by piecewise smooth functions and associated variational problems. *Communications on pure and applied mathematics*, 42(5):577–685, 1989.
19. L. B. Pártay, G. Hantal, P. Jedlovszky, Á. Vincze, and G. Horvai. A new method for determining the interfacial molecules and characterizing the surface roughness in computer simulations. Application to the liquid–vapor interface of water. *Journal of Computational Chemistry*, 29(6):945–956, 2008.
20. S. Plimpton. Fast parallel algorithms for short-range molecular dynamics. *Journal of Computational Physics*, 117(1):1–19, 1995.
21. A. G. Rumpf. Quocmesh software library. Institute for Numerical Simulation, University of Bonn. http://numod.ins.uni-bonn.de/software/quocmesh/.
22. M. Sega, S. S. Kantorovich, P. Jedlovszky, and M. Jorge. The generalized identification of truly interfacial molecules (ITIM) algorithm for nonplanar interfaces. *The Journal of Chemical Physics*, 138(4):044110, 2013.
23. J. A. Sethian. A fast marching level set method for monotonically advancing fronts. *Proceedings of the National Academy of Sciences*, 93(4):1591–1595, 1996.

24. A. Shekhar, K.-i. Nomura, R. K. Kalia, A. Nakano, and P. Vashishta. Nanobubble collapse on a silica surface in water: Billion-atom reactive molecular dynamics simulations. *Phys. Rev. Lett.*, 111:184503, Oct 2013.
25. P. Springer. A scalable, linear-time dynamic cutoff algorithm for molecular simulations of interfacial systems. *arXiv,1502.0323*, 2013.
26. Y. Sun, G. Zheng, C. Mei, E. J. Bohm, J. C. Phillips, L. V. Kalé, and T. R. Jones. Optimizing fine-grained communication in a biomolecular simulation application on cray xk6. In *High Performance Computing, Networking, Storage and Analysis (SC), 2012 International Conference for*, pages 1–11. IEEE, 2012.
27. D. Tameling, P. Springer, P. Bientinesi, and A. E. Ismail. Multilevel summation for dispersion: A linear-time algorithm for $r^{-6}$ potentials. *The Journal of Chemical Physics*, 140(2):024105, 2014.
28. L. Verlet. Computer "Experiments" on Classical Fluids. I. Thermodynamical Properties of Lennard-Jones Molecules. *Phys. Rev.*, 159:98–103, Jul 1967.
29. H. Wang, C. Schütte, and P. Zhang. Error estimate of short-range force calculation in inhomogeneous molecular systems. *Physical Review E*, 86(2):026704, Aug. 2012.
30. C. L. Wennberg, T. Murtola, B. Hess, and E. Lindahl. Lennard-Jones Lattice Summation in Bilayer Simulations Has Critical Effects on Surface Tension and Lipid Properties. *Journal of Chemical Theory and Computation*, pages 3527–3537, July 2013.
31. H. Zhao. Parallel implementations of the fast sweeping method. *Journal of Computational Mathematics*, 25(4), 2007.
32. H.-K. Zhao, S. Osher, B. Merriman, and M. Kang. Implicit and Nonparametric Shape Reconstruction from Unorganized Data Using a Variational Level Set Method. *Computer Vision and Image Understanding*, 80(3):295–314, Dec. 2000.
33. R. A. Zubillaga, A. Labastida, B. Cruz, J. C. Martínez, E. Sánchez, and J. Alejandre. Surface Tension of Organic Liquids Using the OPLS/AA Force field. *Journal of Chemical Theory and Computation*, 9:1611–1615, 2013.